\documentclass[epj]{svjour}
\usepackage{graphicx}
\begin{document}
\title{Sympatric speciation in an age-structured population living on
  a lattice}
\author{A.O. Sousa \thanks{email:sousa@ica1.uni-stuttgart.de}}
\institute{Institute for Computer Applications 1 (ICA1), University of 
Stuttgart, Pfaffenwaldring 27, 70569 Stuttgart, Germany}
\date{Received: date / Revised version: date}
% The correct dates will be entered by Springer
%
\abstract{
A square lattice is introduced into the Penna model for biological
  aging in order to study the evolution of diploid
  sexual populations under certain conditions when one single locus in
  the individual's genome is considered as identifier of species. The
  simulation results show, after several generations, the flourishing
  and coexistence of two separate species in the same environment,
  i.e., one original species splits up into two on the same territory
  (sympatric speciation). As well, the mortalities obtained are in a good
  agreement with the Gompertz law of exponential increase of mortality
  with age. 
 % end of PACS codes
} %end of abstract 
\maketitle
\section{Introduction}
The understanding of species formation - groups of actually or potentially
inter-breeding populations, which are reproductively isolated from
other such groups - is still a fundamental problem in biology
\cite{gen}. Speciation usually occurs when a pre-existing
population is divided into two or more smaller populations by a
geographical barrier, like an island, river, isolated valley, or
mountain range. Once reproductively isolated by the barrier, the gene 
pools in the two populations can diverge due to natural selection, genetic 
drift, or gene flow, and if they sufficiently diverge, then the 
inter-breeding
between the populations will not occur if the barrier is removed. As a 
result, new species have been formed.

In spite of theoretical difficulties to show convincingly how
speciation might occur without physical separation \cite{mayr}, there is an
increasing evidence for the process of sympatric speciation, in which
the origin of two or more species from a single ancestral one 
occurs without geographical isolation \cite{tregenza}. The most 
straightforward scenario for sympatric speciation requires disruptive 
selection
favoring two substantially different phenotypes, followed by 
the elimination of all intermediate phenotypes. In sexual populations, the 
stumbling block preventing sympatric speciation is that mating between 
divergent ecotypes constantly scrambles gene combinations, creating organisms
with intermediate phenotypes. However, this mixing can be prevented
if there is assortative \cite{kirk} instead of random mating, i.e., 
mating of individuals that are phenotypically similar. It can be based on
ecologically important traits such as body size (as in stickle-backs)
\cite{rundle} or on marker traits that co-vary with ecological traits 
(such as
coloration or breeding behavior in cichlids)\cite{wilson}.

The present paper reports on a attempt to address the challenging problem
of sympatric speciation using the widespread Penna bit-string
model \cite{penna,coe} for age-structured populations, which is based on the 
mutation accumulation theory for biological aging. It has successfully 
reproduced many different characteristics of living species, as the 
catastrophic senescence of pacific salmon \cite{salmon}, the inheritance of
longevity \cite{longe} and the evolutionary advantages of sexual
reproduction \cite{sex}, as well as a particular case of sympatric 
speciation 
\cite{spec}.

\section{The model}
\subsection {Model without lattice}
Each individual of the population is represented by a 
``chro\-no\-log\-i\-cal genome'', which consists of two bit-strings 
of 32 bits (32 loci or positions) each, that are read in parallel. 
One string contains the genetic information inherited from the mother 
and the other, from the father. Each position of the bit-strings is 
associated 
to a period of the individual's life, which means that each individual 
can live at most for 32 periods (``years''). Each step of the simulation 
corresponds to reading one new position of all individuals' genomes. 
Genetic defects are represented by bits 1. If an 
individual has two bits 1 at the $i$-th position of both bit-strings 
(homozygote), it will start to suffer the effects of a genetic disease at 
its $i$-th year of life. If the individual is homozygous with 
two bits zero, no disease appears in that age. If the individual 
is heterozygous in that position, it will become sick  
only if that locus is one for which the harmful allele is dominant. The  
dominant loci are randomly chosen at the beginning of
the simulation and remain fixed.  
If the current number of accumulated diseases reaches a 
threshold $T$, the individual dies.

If a female succeeds in surviving until the minimum reproduction
age $R$, it generates $b$ offspring every year until death. The female
randomly chooses a male to mate, the age of which must also be greater
or equal to $R$. The offspring's genome is constructed from the
parent's ones; first the strings of the mother are randomly crossed,
and a female gamete is produced. $M_m$ deleterious mutations are then 
randomly
introduced. The same process occurs with the father's genome (with
$M_f$ mutations), and the union of the two remaining gametes form the
new genome. This procedure is repeated for each of the $b$
offspring. The sex of the baby is randomly chosen, each one with
probability $50\%$. Deleterious mutation means that if a bit 0 is 
randomly chosen in the parent's genome, it is set to 1 in the
offspring genome. However, if a bit already set to 1 is randomly chosen, 
it remains 1 in the offspring genome (no back mutations).

The description given above corresponds to the original sexual version
of the Penna model \cite{book}, in which at every time step each 
individual of the population, independently of its age or current number 
of accumulated diseases, can be killed 
with a probability $V_{t}=1-N_{t}/N_{\rm max}$; $N_{\rm max}$
is the maximum population size (the carrying capacity of the
environment) and $N_{t}$ is the current population size. 
This random time-dependent death, well known as the Verhulst factor, 
is introduced in order 
to avoid the unlimited growth of the population and to take into account 
the dispute for food and space. Since there seems to be no biological
justification for considering random deaths in real populations, as
well as a controversial importance of its role in the Penna model
\cite{verhulst}, in our simulations we do not consider random
deaths. Instead, we adopt a simple lattice dynamics which also  
avoids the exponential increase of the population. 
The details will be presented in the next subsection.

\subsection {Speciation model on a lattice}

In the present case each individual lives on a given site $(i,j)$ of a 
square lattice and, at every time-step, has a probability $p_w$ to move
to the neighboring site that presents the smallest occupation, if
this occupation is also smaller or equal to that of the current
individual's site. We start the simulations randomly distributing one
individual per site on a diluted square lattice. That is, if an
already occupied site is chosen for a new individual, the choice is 
disregarded and another random site is picked out.

At any bit position a diploid individual can have $n$ = 0, 1 or 2 
bits set.   
The process of sympatric speciation is now attempted by defining one
single bit 
position, which we take as position 11, as an identifier of the species.  
Mating occurs only among individuals of the same species (same value 
of $n$ at 
position 11), which means that this locus also defines the mating 
preferences. 
Each able female (with age $\ge R$) with $n$ such bits randomly selects a 
neighboring able male with the same $n$ value to breed. If she succeeds, she 
generates $b$ offspring. Then she chooses at random, again among its 
four neighboring sites, a place to put each baby, according to the
rules below. The newborn dies if it is not possible to find a site 
respecting these rules: 

\noindent 1) The selected site occupation must be $\le$ 1;  

\noindent 2) If the newborn has $n=0$, then it can occupy an empty
site or a site already occupied by a single individual with $n=2$;

\noindent 3) If the newborn has $n=2$, then it can occupy an empty
site or a site already occupied by a single individual with $n=0$;

\noindent 4) If the newborn has $n=1$, it can occupy only an empty site. 

Rules 2 and 3 mean that the $n=0$ and the $n=2$ populations can share 
the same 
habitat, that is, they do not dispute for the same food
resources. Rule 4 means that the $n=1$ population feeds at both niches, 
competing with the other two.  Theses rules replace the random killing 
Verhulst factor pointed out in the previous section.

We start our simulations only with $n=0$ individuals. Due to the 
randomness of 
mutations and crossover, the offspring does not necessarily have 
the same $n$ 
value of the parents. In our model it is exactly this randomness which  
allows the emergence of new species out of the original one. 
These populations coexist in a stable equilibrium but without cross-mating. 

\section{Simulation Results}

The simulation starts with $N_0$ individuals, half males and
half females, and runs for a pre-specified number of time steps, at
the end of which averages are taken over the population(s). The
general parameters of the simulations are:

\noindent  $\bullet$ Minimum age of reproduction $R=8$;

\noindent  $\bullet$ Birth rate $b=3$;

\noindent  $\bullet$ Mutation rate $M=1$ per bit-string (or gamete);

\noindent  $\bullet$ Maximum number of genetic diseases $T=5$; 

\noindent  $\bullet$ Probability to walk $p_{\rm {w}}=1.0$.
 
\begin{figure}[!h]
\begin{center}
\includegraphics[angle=0,scale=0.7]{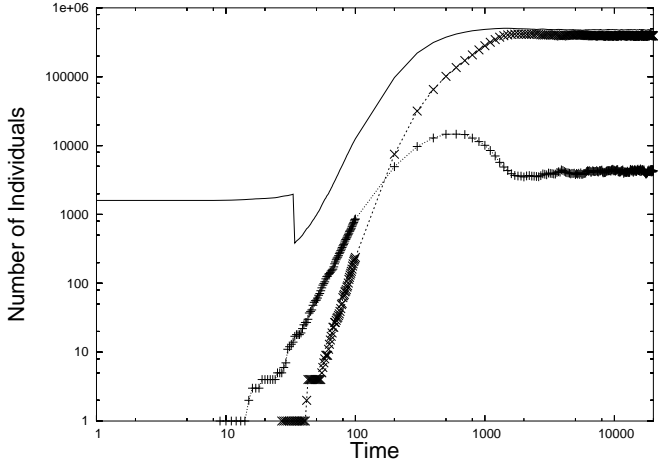}
\end{center}
\caption{Time evolution of $N_0$ (line, original species), $N_1$ (+, mixed
genomes) and $N_2$ (x, new species), for one diploid sexual population
simulated on a $L \times L$ square lattice with $L=800$ and $N_o =
1600$ individuals.}
\label{s_spec1}
\end{figure}

Fig. \ref{s_spec1} shows how the new species $N_2$ emerges, within
about a hundred iterations, from the original species $N_0$. The
intermediate population $N_1$ is only about $0.50 \%$ of the
total population. Since the rule for an individual to move on the
lattice depends only on the existence of a site with an occupation
smaller or equal to that of the current individual's site and is 
completely non-related to the individual species, 
the different species may bunch together at the same site of the
lattice. Then, due to the reproduction rules quoted above, after  
several generations we obtain a great predominance of the two
non-competing species $N_0$ and $N_2$ living at the same geographic 
position (sympatric speciation). Our results with $N_0 = 1600$
are confirmed by larger simulations with $N_0 = 100000$,
and also by larger simulations with $10^6$ time steps.

It must be remarked that the assumptions we make concerning mating
choice and conditions for a newborn to survive were adopted in order
to capture some features of field observations and laboratory
experiments of species which seem to speciate via disruptive
selection on habitat/food preferences and assortative mating. 

 For instance, in a series of
papers Rice and Salt \cite{rice} presented experimental evidence for  
the possibility of sympatric speciation in {\it Drosophila
melanogaster}. They started from the premise that whenever organisms sort
themselves into the environment first and then mate locally, individuals with
the same habitat preferences will necessarily mate assortatively.
Others examples of sympatric speciation can be found for canids 
\cite{canids},
lizards \cite{lizards} and pandas \cite{panda}. In this latter example,
the Giant Panda ({\it Ailuropoda melanoleuca}) and the Red Panda
({\it Ailurus fulgens}) are vegetarian carnivores that specialize in
eating bamboo in Sichuan Province, China. The two species share the
same habitats and bamboo plants. Both pandas feed on the same species 
of bamboo, but specialize in eating different parts of the bamboo
plant. The Giant Panda feeds more frequently on bamboo stems, while
the Red Panda feeds more frequently on bamboo leaves \cite{panda}.
In our simulation, disruptive selection explicitly arises from
competition for a single resource (a potentially more common
ecological situation). In this way, we may imagine, for instance, that
the original population $n=0$ is vegetarian, and the second population
$n=2$ emerging out of it consists of carnivores (thus, there is no
competition between the two different populations). However, since the
individuals with $n=1$ feed from the same resources of both
populations ($n=0$ and $n=2$), this competition for food   
reduces its abundance in the system and, 
combined with assortative mating, leads to evolutionary branching.

The situation that better fits our simulations occurs in the Australian 
Fogg Dam Nature Reserve, where data have been collected \cite{JTE} from 
three 
different snake species: water phytons ({\it Liasis fuscus, Pythonidae}), 
keelbacks ({\it Tro\-pi\-donophis mairii, Colubridae}) and slatey-grey 
snakes 
({\it Stegonotus cucullatus, Colubridae}). All are non-venous, 
ovip\-a\-rous and active foragers, but they differ considerably in
body sizes and dietary habits. Water phytons feed almost exclusively
on a single species of native 
rodent; keelbacks feed primarily on frogs and slatey-grey snakes have 
extremely broad diets (reptile eggs, frogs, small mammals and lizards). 
According to Ref.\cite{JTE}, the population of slatey-grey snakes is 
smaller  
than the other two during the whole year. Particularly from April to 
May (when neither the rats nor the frogs are in their peaks    
of abundance),  the water phyton and the keelbacks populations are almost 
of the same size while the slatey-grey snakes population size is around 
$1/7$ of this value. 
   
\begin{figure}
\begin{center}
\includegraphics[angle=0,scale=0.7]{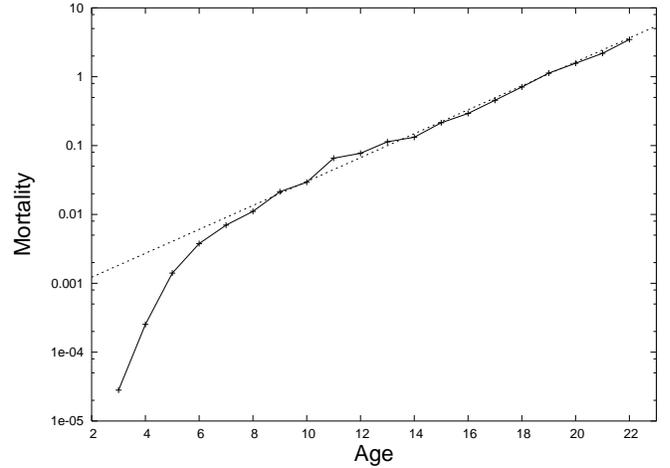}
\end{center}
\caption{Mortality as a function of age averaged
  over $20$ diploid sexual populations simulated on a square lattice
  with $L=200$ and $N_0 = 800$ individuals. The dashed line corresponds
  to the fit $q(\rm{age})=0.00055\exp(0.4*\rm{age})$ in this
  semilogarithmic plot.}
\label{s_death}
\end{figure}

To check the main responsible element for the observed speciation in our
model, we have also performed several simulations considering all
possible permutations of the four new aspects introduced into the
usual aging model here studied:
 
{\bf 1)} the kind of species in the beginning of the simulation 
(only $N_0$ or $N_1$ or $N_2$); 

{\bf 2)} Mating preferences: The mating occurs only among individuals 
of the same species or it occurs without taking into account the 
individual's species (randomly).

{\bf 3)} Back mutation: A reversal process whereby a gene that has 
undergone mutation returns to its previous state, i.e., if the randomly 
chosen position for introducing mutation at birth is the bit identifier 
of the species (position $11$), with a small probability $p_{bm}$, a 
bit set to $1$ in the parent's genome, it is set to $0$ in the offspring 
genome. However, if this bit is already set to $0$, it remains $0$ in 
the offspring genome. For all the others positions in the genetic 
strand only the possibility of harmful mutation is considered. 

{\bf 4)} With or without the coexistence rule: the newborn with $n=0$ 
($n=2$) can occupy an empty site or a site already occupied by a single 
individual with $n=2$ ($n=0$), or they can born if and only if there is 
an empty site to be placed.

Additionally, for all the cases, we also analyzed if there would be any 
finite size effects related to the number of sites which the newborn 
can be placed: the newborn can be randomly placed only on the 
nearest-neigh\-bor mother's site or in any site of the lattice.

If the simulation starts only with individuals $N_0$ or only with 
individuals $N_1$, and only harmful mutations are introduced, the
speciation is observed only if the coexistence rule is considered. 
However, for the case starting with only individuals $N_2$, the time
evolution of the population presents only individuals $N_2$, since if 
only deleterious mutations are considered one bit set to $1$ does not 
change to one set to $0$. That's why in this case, the speciation 
could be observed only when the back mutations with 
$p_{bm}=5\times10^{-5}$ were allowed. Furthermore, in all cases 
starting only with species $N_2$, the newborns suffer back-mutations 
with a small probability $p_{bm}$.

\begin{figure}[!h]
\begin{center}
\includegraphics[angle=0,scale=0.7]{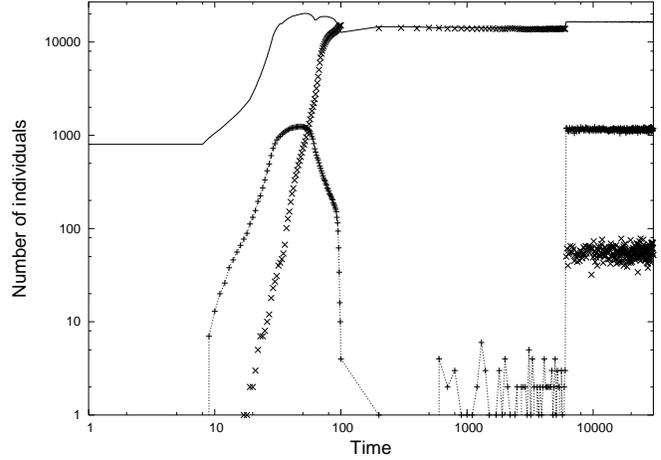}
\end{center}
\caption{Time evolution of $N_0$ (line, original species), 
$N_1$ (+, mixed ge\-nomes) and $N_2$ (x, new species), for one 
diploid sexual population simulated on a $L \times L$ square 
lattice with $L=800$ and $N_0 = 1600$ individuals, when two different 
strategies for the placement of newborns are considered. For 
$t < t_1$, newborns can be placed on empty sites and those ones with
$n=0$ and $n=2$ can share the same lattice site, when $t \ge t_1$ 
newborns can born if and only if there is any empty site on the
lattice. $t_{1}=6000$.}
\label{s_spec3}
\end{figure}

If there is no mating preference, which means that individuals from 
different species can mate with one another, then the speciation
occurs, however the coexistence rule must be considered, as mentioned 
before. The results obtained when the newborns are allowed to occupy
only the nearest-neigh\-bor mother's sites and those ones when it
could also occupy any lattice site show qualitatively the same
behavior. It has been also noticed that the vanishing of intermediate 
individuals ($N_1$) (Fig. \ref{s_spec1}) was not caused by the restriction 
that their offspring could be born only if there is an empty site in 
the nearest-neighbor mothers' site, since simulations taking into account 
the nearest-neighbor, the next-nearest-neighbor mother's site and any 
empty site of the lattice (independently of its neighborhood) did not 
avoid their extinction, which, in fact, was observed to be due to the 
competition with the others two species ($N_0$ and $N_2$) for the same 
resources, that is imposed through the prohibition of newborns $N_1$ 
to born on sites which are already occupied. This conclusion could be 
reinforced with the findings of the following investigation: 
During a certain timestep $t < t_1$, the population evolves
considering the possibility to put newborns on empty sites and also
the coexistence rule. For $t \ge t_1$, when the population size has reached 
the equilibrium (constant in time), the coexistence rule is disregard and 
only one condition for the survival of the offspring is hold: the
newborn will be born only if there exists any empty lattice site 
(not only in the mother's neighborhood) for it to be placed on. As we
can see from the Figure \ref{s_spec3}, when $t < t_1$, the sympatric 
speciation occurs and the results are similar to one showed in Figure 
\ref{s_spec1}, $(N_{0} \simeq N_{2})>>N_{1}$. However, for 
$t \ge t_1$, since the population size of the mixed species is larger 
than those with $n=2$ ($N_0 > N_1 > N_2$), it is not possible anymore 
to affirm the existence of speciation, even the coexistence of two 
separate species in the same environment (site). Based on this result, 
we can conclude that the sympatric speciation obtained 
in our simulations is essentially caused by non-existence of competition 
between the newborns with $N_0$ and $N_2$ and not by the finite size 
effects related to the neighborhood considered to place the newborns 
$N_1$.

If the females mate with males from any species randomly, the coexistence rule 
is the only condition required for ocurring speciation, excepting when 
the simulation starts with species $N_2$ we must also take into account 
the back-mu\-ta\-tions, for the reason previously mentioned. From 
this result, we can conclude that the mating preference condition is 
not necessary to be imposed in order to obtain sympatric speciation in 
our model, but the species's ecological behaviour introduced through
the coexistence rule. This conclusion is more reinforced with the
results from all the simulations discussed before in which the
coexistence rule always must be considered to lead the population to
speciation. 

As a final study, we examine our populations mortalities. In 1825, based on
observed death and population records of people in England, Sweden,
and France between ages $20$ and $60$ in the nineteenth century, the
British actuary Benjamin Gompertz derived a simple formula
describing the exponential increase in death rates between sexual
maturity and extreme old ages \cite{gompertz}. This formula, 
$q(\rm{age})=\rm{A} \times \exp(\rm{b} \times \rm{age})$,  
is commonly referred to as the Gompertz's law of mortality. 
As Fig. \ref{s_death} shows, our results for the mortality above 
the minimum reproduction age $R=8$, are in a good agreement with 
the Gompertz law. 

\section{Conclusions}
In conclusion, the results presented here are based on a very simple 
assumption that a single locus in the individual's genome determines 
the ecological behavior of the individual and identifies its species,
in the framework of the Penna bit-string model. Despite its
simplicity and limited applicability, our results clearly show the emergence
of sympatric speciation in diploid sexual age-structured populations 
of individuals that are distributed on a square lattice.
The introduction into the model of an specific gene responsible for 
the ecological behaviour of the individual: individuals with $n=1$ and
$n=2$ do not compete for food resources, so they can bunch together at
the same lattice site, tailored to reproduce recent observations of 
existing species \cite{greenberg} that have led to the suggestion 
that ecological adaptation is the driving force behind divergence 
of populations leading to speciation: a gene, {\it desaturase2}, of 
{\it Drosophila melanogaster}, which confers resistence to cold 
as well as susceptibility to starvation, underlies a pheromonal 
difference and contributes to the reproductive isolation between 
some Drosophila species: the Zimbabwe and Cosmopolitan races.

Since the speciation discussed in this paper is triggered by 
interaction or competion between organisms, and not merely by
mutation, the process is not so much random as deterministic.
In fact, the speciation process occurs irrespectively of the adopted 
random number in the simulation, as well without assuming mating 
preference. According to our scenario, the coexistence of the two
groups ($N_0$ and $N_2$) is the main mechanism that often leads the
population to speciation. It should be also noted that the change 
in genotypes occur within few generations. The speed of genetic change, 
of course, depends on the mutation rate, but the present mechanism is 
found to work even for any smaller mutation rate (say $T \le 5 $).

The relationship between our model and previous phys\-ics models, like 
directed percolation or reaction-diffusion problems, as suggested 
by an anonymous referee, unfortunately remains to be elucidated, 
even though some relations to reaction-diffusion systems can be e
stablished, {\it a  priori}. By the way, further research is needed 
to explore that question, which it is not the goal and the context 
of the present paper.

\begin{acknowledgement}
I would like to thank Dietrich Stauffer and S. Moss de Oliveira for
very important discussions and a critical reading of the
manuscript. I appreciate a discussion with Peter Grassberger, who had
denied a strong relation of the present model to directed percolation and
pointed out some possible relations to reaction-diffusion systems; and
the helpful suggestions of the anonymous referees. This work was
supported by a grant from Alexander vonHumboldt Foundation.
\end{acknowledgement}

\end{document}